\documentclass[%
 aip,
 amsmath,amssymb,
reprint,%
]{revtex4-1}

\usepackage{amsmath}
\usepackage{xspace}
\usepackage{multirow}
\usepackage{latexsym}
\usepackage{braket}
\usepackage{cleveref}
\usepackage{threeparttable}
\usepackage{enumerate}
\usepackage{mathrsfs}
\usepackage{enumitem}
\usepackage{color}
\usepackage[version=3]{mhchem}
\usepackage{caption,setspace}
\usepackage{graphicx}
\usepackage{subfigure}
\usepackage{threeparttable}

\usepackage{array}
\newcolumntype{L}[1]{>{\raggedright\let\newline\\\arraybackslash\hspace{0pt}}m{#1}}
\newcolumntype{C}[1]{>{\centering\let\newline\\\arraybackslash\hspace{0pt}}m{#1}}
\newcolumntype{R}[1]{>{\raggedleft\let\newline\\\arraybackslash\hspace{0pt}}m{#1}}

\crefname{figure}{Figure}{Figures}
\crefname{table}{Table}{Tables}
\crefname{equation}{Eq.}{Eqs.}
\crefname{section}{Section}{Sections}

\newcommand*{\mae}{$\Delta_{\mathrm{MAE}}$\xspace}
\newcommand*{\std}{$\Delta_{\mathrm{STD}}$\xspace}
\newcommand*{\eh}{\ensuremath{E_\mathrm{h}}\xspace}

\begin{document}

\author{Samragni~Banerjee}
\affiliation{Department of Chemistry and Biochemistry, The Ohio State University, Columbus, Ohio 43210, USA}

\author{Alexander~Yu.~Sokolov}
\email{sokolov.8@osu.edu}
\affiliation{Department of Chemistry and Biochemistry, The Ohio State University, Columbus, Ohio 43210, USA}

\title{Third-order algebraic diagrammatic construction theory for electron attachment and ionization energies: Conventional and Green's function implementation}

\begin{abstract}

We present implementation of second- and third-order algebraic diagrammatic construction theory for efficient and accurate computations of molecular electron affinities (EA), ionization potentials (IP), and densities of states (EA-/IP-ADC(n), n = 2, 3). Our work utilizes the non-Dyson formulation of ADC for the single-particle propagator and reports working equations and benchmark results for the EA-ADC(2) and EA-ADC(3) approximations. We describe two algorithms for solving EA-/IP-ADC equations: (i) conventional algorithm that uses iterative diagonalization techniques to compute low-energy EA, IP, and density of states, and (ii) Green's function algorithm (GF-ADC) that solves a system of linear equations to compute density of states directly for a specified spectral region. To assess accuracy of EA-ADC(2) and EA-ADC(3), we benchmark their performance for a set of atoms, small molecules, and five DNA/RNA nucleobases. As our next step, we demonstrate efficiency of our GF-ADC implementation by computing core-level $K$-, $L$-, and $M$-shell ionization energies of a zinc atom without introducing core-valence separation approximation. Finally, we use EA- and IP-ADC methods to compute band gaps of equally-spaced hydrogen chains H$_n$ with $n$ up to 150, providing their estimates near thermodynamic limit. Our results demonstrate that EA-/IP-ADC(n) (n = 2, 3) methods are efficient and accurate alternatives to widely used electronic structure methods for simulations of electron attachment and ionization properties.

\end{abstract}

\titlepage

\maketitle

\section{Introduction}
\label{sec:intro}

Accurate computations of electron affinities (EA) and ionization potentials (IP) are important for predicting properties of molecules and materials, such as redox potentials, band gaps, and photoelectron 
spectra.\cite{Cederbaum:1975p290,VonNiessen:1984p57,Ortiz:2012p123,Hirata:2015p1595,Hirata:2017p044108}
For molecular systems, a variety of theoretical approaches for computing EA and IP have been developed, ranging from affordable density functional theory (DFT) and time-dependent many-body perturbation theory (MBPT)\cite{Hedin:1965p796,Faleev:2004p126406,vanSchilfgaarde:2006p226402,Neaton:2006p216405,Samsonidze:2011p186404,vanSetten:2013p232,Reining:2017pe1344} approximations 
to highly accurate coupled cluster (CC) methods in their state-specific or equation-of-motion (EOM) 
formulations.\cite{Crawford:2000p33,Shavitt:2009,Nooijen:1992p55,Nooijen:1993p15,Nooijen:1995p1681,Kowalski:2014p094102,BhaskaranNair:2016p144101,McClain:2016p235139,McClain:2017p1209,Lange:2018p4224,Peng:2018p4335}
Although materials simulations have been traditionally dominated by DFT and MBPT, recent methodological advances and increase in computer power have enabled computations of band structures of three-dimensional semiconductors using equation-of-motion coupled cluster theory (EA/IP-EOM-CC), in a good agreement with experimental 
results.\cite{McClain:2016p235139,McClain:2017p1209} 
Despite these initial successes, simulations of EA and IP of solids and large molecular systems using accurate {\it ab initio} methods such as EOM-CC are far from routine, primarily due to their high computational cost.

A less expensive alternative to CC theory for simulating excited-state properties of molecules is algebraic diagrammatic construction theory 
(ADC).\cite{Schirmer:1982p2395,Schirmer:1983p1237,Schirmer:1991p4647,Mertins:1996p2140,Schirmer:2004p11449,Dreuw:2014p82} 
The ADC approximations are derived from a perturbative expansion of propagators, poles and residues of which provide information about excitation energies and transition intensities. Although originally formulated within the framework of time-dependent MBPT, ADC was later rederived in the context of time-independent perturbation theory using the intermediate-state representation\cite{Schirmer:1991p4647,Mertins:1996p2140,Schirmer:2004p11449} and effective Liouvillean\cite{Mukherjee:1989p257} approaches. More recently, it has been demonstrated that ADC emerges as an approximation in the linear-response unitary CC (UCC)\cite{Kats:2011p062503,Walz:2012p052519} and self-consistent UCC-based polarization propagator\cite{Liu:2018p244110} theories. Although computational scaling of low-order ADC approximations (e.g., ADC(3)) is similar to that of traditional CC methods with single and double excitations (CCSD), computation of excitation energies and transition properties using the ADC methods is more efficient due to the non-iterative and Hermitian nature of the ADC equations. 

The ADC approximations have been applied to simulations of a variety of excited-state properties, including optical excitation,\cite{Schirmer:1982p2395,Starcke:2009p024104,Harbach:2014p064113} ionization,\cite{Schirmer:1983p1237,Schirmer:1998p4734,Trofimov:2005p144115,Angonoa:1987p6789,Schirmer:2001p10621,Thiel:2003p2088,Dempwolff:2019p064108} core\cite{Barth:1985p867,Wenzel:2014p1900} and two-photon\cite{Knippenberg:2012p064107} absorption. For computations of IP and EA, the ADC methods were originally formulated\cite{Schirmer:1983p1237} based on the perturbative self-energy expansion of the Dyson equation (so-called Dyson ADC framework), which couples IP ($N-1$) and EA ($N+1$) components of the one-electron propagator. In 1998, Schirmer and co-workers proposed the non-Dyson ADC formulation,\cite{Schirmer:1998p4734} which decouples the $N\pm1$ parts of the propagator and allows for independent calculations of IP and EA. Since then, several implementations of the single-reference non-Dyson IP-ADC methods up to third order in perturbation theory (IP-ADC(n), n = 2, 3) have been developed and applied to closed- and open-shell systems.\cite{Trofimov:2005p144115,Dempwolff:2019p064108} For computations of IP of strongly correlated systems, a multi-reference formulation of IP-ADC(2) has been recently implemented in our group.\cite{Chatterjee:2019p5908} Two implementations of the non-Dyson EA-ADC(n) (n = 2, 3) approximations have been reported,\cite{Trofimov:2011p77,Schneider:2015} although performance of these methods has not been documented thoroughly in the literature. 

In this work, we present a new implementation of the non-Dyson EA- and IP-ADC methods up to third order in perturbation theory (EA-/IP-ADC(n), n = 2, 3) using two algorithms. In the first algorithm, we take the conventional route and use iterative diagonalization techniques to compute ionization and electron attachment energies. In the second algorithm, we bypass the diagonalization of the ADC effective Hamiltonian matrix and compute the spectral function directly by solving a system of linear equations at each frequency of the external field, in a fashion similar to that of Green's function coupled cluster 
methods.\cite{McClain:2016p235139,BhaskaranNair:2016p144101,Peng:2018p4335} 
We derive working equations for the EA-/IP-ADC(n) (n = 2, 3) methods using the formalism of effective Liouvillean theory, present their implementation, and benchmark results of EA-ADC(2) and EA-ADC(3) for a variety of closed- and open-shell systems, including atoms, small molecules, and five DNA/RNA nucleobases (\cref{sec:results:ea_atoms_small_molecules,sec:results:nucleobases}). We also demonstrate capabilities of our Green's function ADC implementation by computing core ionization energies of a zinc atom without introducing core-valence approximation (\cref{sec:results:zinc}). Finally, we combine our EA- and IP-ADC implementations to compute band gaps of the equally-spaced hydrogen chains H$_{n}$ for $n$ up to 150 (\cref{sec:results:hydrogen_chains}) and present our conclusions (\cref{sec:conclusions}).

\section{Theory}
\label{sec:theory}
\subsection{General overview of ADC}
The algebraic diagrammatic construction (ADC) theory has a close connection with the propagator theory,\cite{Fetter2003,Dickhoff2008} where the central object of interest is a propagator of the many-body system. The general expression for a propagator that describes electron attachment (EA) and removal (IP) processes is given by the retarded single-particle Green's function:
\begin{align}
G_{pq}(\omega) &= G_{pq}^{+}(\omega) + G_{pq}^{-}(\omega)\notag\\
               &=\langle\Psi_{0}^{N}|a_{p} (\omega - H + E_{0}^{N})^{-1}a_{q}^{\dagger}|\Psi_{0}^{N}\rangle\notag\\
               &+ \langle\Psi_{0}^{N}|a_{q}^{\dagger} (\omega + H - E_{0}^{N})^{-1}a_{p}|\Psi_{0}^{N}\rangle\label{eq:1}
\end{align}
where $G_{pq}^{+}(\omega)$ and $G_{pq}^{-}(\omega)$ are the forward (EA) and backward (IP) components of $G_{pq}(\omega)$, \textit{H} is the electronic Hamiltonian, $|\Psi_{0}^{N}\big\rangle$ and \textit{E}$_{0}^{N}$ are the ground-state \textit{N}-electron wavefunction and energy, respectively. The frequency can be expressed as $\omega$ $\equiv$ $\omega' + i\eta$, where $\omega'$ is the real part of $\omega$ and $i\eta$ is an infinitesimal imaginary number. The operators $a_{p}$ and $a_{p}^{\dag}$ are the usual one-electron annihilation and creation operators, respectively. 

The propagator in \cref{eq:1} can be expressed using the Lehmann (or spectral) representation:
\begin{align}
 G_{pq}(\omega) &= \sum_{n}\frac{\langle\Psi_{0}^{N}|a_{p}|\Psi_{n}^{N+1}\rangle \langle\Psi_{n}^{N+1}|a_{q}^{\dagger}|\Psi_{0}^{N}\rangle}{\omega - E_{n}^{N+1}+ E_{0}^{N}}\notag\\
 &+  \sum_{n}\frac{\langle\Psi_{0}^{N}|a_{q}^{\dagger}|\Psi_{n}^{N-1}\rangle \langle\Psi_{n}^{N-1}|a_{p}|\Psi_{0}^{N}\rangle}{\omega + E_{n}^{N-1} - E_{0}^{N}},\label{eq:2}
 \end{align}
Here, the resolution of identity is carried out over all exact eigenstates $\ket{\Psi_{n}^{N+1}}$ ($\ket{\Psi_{n}^{N-1}}$) of the electron-attached (-ionized) system with energies $E_{n}^{N+1}$ ($E_{n}^{N-1}$).
In \cref{eq:2}, the poles of $G_{pq}(\omega)$ provide information about vertical electron attachment ($\omega_{n} = E_{n}^{N+1} - E_{0}^{N}$) and ionization ($\omega_{n} = E_{0}^{N} - E_{n}^{N - 1}$) energies, while the residues (e.g., $\langle\Psi_{0}^{N}|a_{p}|\Psi_{n}^{N+1}\rangle \langle\Psi_{n}^{N+1}|a_{q}^{\dagger}|\Psi_{0}^{N}\rangle$) describe probabilities of the corresponding transitions. For each component of the propagator, the Lehmann representation can be compactly written in a matrix form as:
\begin{align}
    \textbf{G}_{\pm}(\omega) = \mathbf{\tilde{X}}_{\pm}(\omega - \boldsymbol{\tilde{\Omega}}_{\pm})^{-1}\mathbf{\tilde{X}}_{\pm}^{\dagger}\label{eq:3}
\end{align}
where $\boldsymbol{\tilde{\Omega}}_{\pm}$ are the diagonal matrices of vertical attachment/ionization energies ($\omega_{n}$) and $\mathbf{\tilde{X}}_{\pm}$ are matrices of the transition amplitudes with elements $\tilde{X}_{+pn} = \langle\Psi_{0}^{N}|a_{p}|\Psi_{n}^{N+1}\rangle$ and $\tilde{X}_{-qn} = \langle\Psi_{0}^{N}|a_{q}^{\dagger}|\Psi_{n}^{N-1}\rangle$. 

The single-particle Green's function in \cref{eq:1,eq:2,eq:3} is exact if expressed using the exact eigenstates of the system. However, computing the exact propagator is computationally very expensive and many approximations have been introduced to reduce the computational cost.\cite{Hedin:1965p796,Faleev:2004p126406,vanSchilfgaarde:2006p226402,Neaton:2006p216405,Samsonidze:2011p186404,vanSetten:2013p232,Reining:2017pe1344,Nooijen:1992p55,Nooijen:1993p15,Nooijen:1995p1681,Kowalski:2014p094102,BhaskaranNair:2016p144101,McClain:2016p235139,McClain:2017p1209,Lange:2018p4224,Peng:2018p4335} 
In the ADC theory,\cite{Schirmer:1982p2395,Schirmer:1983p1237,Schirmer:1991p4647,Mertins:1996p2140,Schirmer:2004p11449,Dreuw:2014p82} \cref{eq:3} is rewritten in a basis of ($N+1$)- and ($N-1$)-electron states where the matrix representation of the Hamiltonian $H$ is no longer diagonal and each component of the Green's function can be written as:
\begin{align}
    \textbf{G}_{\pm}(\omega) = \textbf{T}_{\pm}(\omega\textbf{S}_{\pm}-\textbf{M}_{\pm})^{-1}\textbf{T}_{\pm}^{\dagger}\label{eq:4}
\end{align}
Here, $\textbf{M}_{\pm}$ and $\textbf{T}_{\pm}$ are known as effective Hamiltonian and transition moments matrices, respectively. For generality, we assume that the basis of $N\pm1$ states is non-orthogonal with the overlap matrix $\textbf{S}_{\pm}$. We note that in the original ADC theory\cite{Schirmer:1982p2395,Schirmer:1983p1237,Schirmer:1991p4647,Mertins:1996p2140,Schirmer:2004p11449,Dreuw:2014p82} \cref{eq:4} is usually expressed in the basis of orthogonalized intermediate states, where $\textbf{S}_{\pm}$ = $\mathbf{1}$.

The non-Dyson ADC scheme\cite{Schirmer:1998p4734} evaluates the propagators $\textbf{G}_{\pm}(\omega)$ by expanding each matrix in \cref{eq:4} in a perturbative series and truncating the expansions at the $n$-th order:
\begin{align}
    \textbf{M}_{\pm} &\approx \textbf{M}^{(0)}_{\pm} + \textbf{M}^{(1)}_{\pm} + ...+ \textbf{M}^{(n)}_{\pm}\label{eq:5}\\
     \textbf{S}_{\pm} &\approx \textbf{S}^{(0)}_{\pm} + \textbf{S}^{(1)}_{\pm} + ...+ \textbf{S}^{(n)}_{\pm}\label{eq:6}\\
   \textbf{T}_{\pm} &\approx \textbf{T}^{(0)}_{\pm} + \textbf{T}^{(1)}_{\pm} + ...+ \textbf{T}^{(n)}_{\pm}\label{eq:7}
\end{align}
Plugging \cref{eq:5,eq:6,eq:7} into \cref{eq:4} leads to equations of the $n$-th order ADC approximations for electron attachment and ionization energies (EA-/IP-ADC(n)). The vertical electron attachment/ionization energies can be computed by solving the generalized eigenvalue problem:
\begin{align}
    \textbf{M}_{\pm}\textbf{Y}_{\pm}=\textbf{S}_{\pm}\textbf{Y}_{\pm}\boldsymbol{\Omega}_{\pm}\label{eq:8}
\end{align}
where $\boldsymbol\Omega_{\pm}$ is a diagonal matrix of eigenvalues and \textbf{Y}$_{\pm}$ are the eigenvectors that can be used to compute spectroscopic amplitudes 
\begin{align}
    \textbf{X}_{\pm}=\textbf{T}_{\pm}\textbf{S}_{\pm}^{-\frac{1}{2}}\textbf{Y}_{\pm}\label{eq:9}
\end{align}
which provide information about transition intensities. Using $\boldsymbol{\Omega}_{\pm}$ and $\mathbf{X}_{\pm}$ allows to compute the ADC propagator and density of states
\begin{align}
	\label{eq:g_adc}
	\mathbf{G}_{\pm}(\omega) &= \mathbf{X}_{\pm} \left(\omega - \boldsymbol{\Omega}_{\pm}\right)^{-1}  \mathbf{X}_{\pm}^\dag \\
	\label{eq:spec_function}
	A(\omega) &= -\frac{1}{\pi} \mathrm{Im} \left[ \mathrm{Tr} \, \mathbf{G}_{\pm}(\omega) \right]
\end{align}
A widely used approach for deriving the ADC equations is via the intermediate state representation (ISR) formalism.\cite{Schirmer:1991p4647,Mertins:1996p2140,Schirmer:2004p11449} In the next section, we present an alternative derivation of the EA-/IP-ADC equations using the formalism of the effective Liouvillean theory.\cite{Mukherjee:1989p257,Sokolov:2018p204113}

\subsection{Non-Dyson EA- and IP-ADC from the effective Liouvillean theory}
To derive EA-/IP-ADC equations using the effective Liouvillean framework, we start by expressing the exact ground-state wavefunction using a unitary cluster expansion:
\begin{align}
|\Psi_{0}^{N}\rangle = e^{T - T^{\dagger}}|\Phi\rangle\label{eq:10}
\end{align}
where $|\Phi\rangle$ is a reference wavefunction. In this work, we assume that $|\Phi\rangle$ is a single Slater determinant and refer interested readers to Ref.\@ \citenum{Sokolov:2018p204113} for the generalization of the effective Liouvillean theory to multi-determinant wavefunctions. The operator $T$ is an excitation operator that generates all possible excitations from the occupied to virtual (external) orbitals labeled using the $i,j,k,l,\ldots$ and $a,b,c,d,\ldots$ indices, respectively:
\begin{align}
    T = \sum_m^N T_{m}, \quad T_{m} = \frac{1}{(m!)^{2}}\sum_{ijab\ldots} t_{ij\ldots}^{ab\ldots}a_{a}^{\dagger}a_{b}^{\dagger}\ldots a_{j}a_{i}\label{eq:11}
\end{align}

We now define the electronic Hamiltonian $H$ 
\begin{align}
 H = \sum_{pq}h_{p}^{q}a_{p}^{\dagger}a_{q} + \frac{1}{4}\sum_{pqrs}v_{pq}^{rs}a_{p}^{\dagger}a_{q}^{\dagger}a_{s}a_{r}\label{eq:12}
\end{align}
where $h_{p}^{q}=\langle p|h|q\rangle$ and $v_{pq}^{rs} = \langle pq||rs\rangle$ are the one-electron and two-electron integrals, respectively, with the $p,q,r,s$ indices running over the entire spin-orbital basis set. Normal-ordering the creation and annihilation operators with respect to $|\Phi\rangle$, the Hamiltonian can be expressed as:
\begin{align}
    H = E_{0} + \sum_{pq}f_{p}^{q}\{a_{p}^{\dagger}a_{q}\} + \frac{1}{4}\sum_{pqrs}v_{pq}^{rs}\{a_{p}^{\dagger}a_{q}^{\dagger}a_{s}a_{r}\}\label{eq:13}
\end{align}
where $E_{0} = \langle\Phi|H|\Phi\rangle$ and $f_{p}^{q} = h_{p}^{q} + \sum_{i}^{occ}v_{pi}^{qi}$ are the reference energy and canonical Fock matrix, respectively, and the normal ordering of the creation and annihilation operators is denoted by the curly braces $\{\ldots\}$. For convenience, we will assume that the Fock matrix is diagonal, i.e.\@ $f_{p}^{q} = \delta_{p}^{q} \epsilon_p$.

To define the ADC perturbative series, we partition the Hamiltonian into the zeroth-order part 
\begin{align}
    H^{(0)} = E_{0} + \sum_{p} \epsilon_{p}\{a_{p}^{\dagger}a_{p}\}\label{eq:14}
\end{align}
and a perturbation $V = H -H^{(0)}$. This leads to a perturbative expansion for the ground-state wavefunction and the propagator 
\begin{align}
    |\Psi_0^N\rangle &= e^{T^{(0)} - T^{(0)\dagger}+T^{(1)} - T^{(1)\dagger}+\ldots+T^{(n)} - T^{(n)\dagger}+\ldots}|\Phi\rangle\label{eq:15} \\
    \textbf{G$_{\pm}(\omega)$} &= \textbf{G$_{\pm}^{(0)}(\omega)$} +\textbf{G$_{\pm}^{(1)}(\omega)$} + ... + \textbf{G$_{\pm}^{(n)}(\omega)$} + ... \label{eq:16}
\end{align}
Truncating this expansion at the $n$-th order yields working equations for the matrix elements of $\textbf{M}_{\pm}$, $\textbf{S}_{\pm}$, and $\textbf{T}_{\pm}$ in  \cref{eq:4,eq:5,eq:6,eq:7}.

For the forward propagator $\textbf{G}_{+}(\omega)$ describing electron attachment (EA-ADC), expressions for the $n$-th-order contributions to the ADC matrices are given as:
\begin{align}
    M_{+\mu\nu}^{(n)} &= \sum_{klm}^{k+l+m=n} \langle \Phi|[h_{+\mu}^{(k)},[\tilde{H}^{(l)},h_{+\nu}^{(m)\dagger}]]_{+}|\Phi\rangle \label{eq:17} \\
    S_{+\mu\nu}^{(n)} &= \sum_{kl}^{k+l=n} \langle \Phi|[h_{+\mu}^{(k)},h_{+\nu}^{(l)\dagger}]_{+}|\Phi\rangle \label{eq:18}\\
    T_{+p\nu}^{(n)} &= \sum_{kl}^{k+l=n} \langle \Phi|[\tilde{a}_{p}^{(k)},h_{+\nu}^{(l)\dagger}]_{+}|\Phi\rangle \label{eq:19}
\end{align}
where $[\ldots]$ and $[\ldots]_{+}$ denote commutator and anti-commutator, respectively. The operators $\tilde{H}^{(k)}$ and $\tilde{a}_{p}^{(k)}$ are the $k$-th-order contributions to the effective Hamiltonian $\tilde{{H}}=e^{-(T-T^{\dagger})}He^{(T-T^{\dagger})}$ and observable $\tilde{a}_{p}=e^{-(T-T^{\dagger})}a_{p}e^{(T-T^{\dagger})}$ operators. These contributions are obtained by expanding $\tilde{H}$ and $\tilde{a}_{p}$ using the Baker--Campbell--Hausdorff (BCH) formula and collecting terms at the $k$-th order:
\begin{align}
    \tilde{H} &= H^{(0)} + V + [H^{(0)},T^{(1)} - T^{(1){\dagger}}] + [H^{(0)},T^{(2)} - T^{(2){\dagger}}]\notag   \\
    &+\frac{1}{2!}[V+(V + [H^{(0)},T^{(1)} - T^{(1){\dagger}}]),T^{(1)}-T^{(1){\dagger}}]+...\label{eq:20}\\ 
    \tilde{a}_p &= a_p + [a_p,T^{(1)}-T^{(1){\dagger}}]+ [a_p,T^{(2)}-T^{(2){\dagger}}]\notag\\
    &+\frac{1}{2!}[[a_p,T^{(1)}-T^{(1){\dagger}}],T^{(1)}-T^{(1){\dagger}}]+...\label{eq:21}
\end{align}
The operators $h_{+\mu}^{(k)\dagger}$ in \cref{eq:17,eq:18,eq:19} are electron attachment operators that are used to construct a set of basis states $\ket{\Psi^{(k)}_{+\mu}} = h_{+\mu}^{(k)\dagger}\ket{\Phi}$ necessary for representing the eigenstates of the $(N+1)$-electron system in \cref{eq:8}. Up to first order in perturbation theory, these operators have the form $h_{+\mu}^{(0)\dagger} = a_{a}^{\dagger}$ and ${h}_{+\mu}^{(1)\dagger} =a_b^{\dagger}a_a^{\dagger}a_{i}$.

For electron ionization (IP-ADC) described by the backward propagator $\textbf{G}_{-}(\omega)$, the $n$-th-order ADC matrices are expressed in a similar form:
\begin{align}
    M_{-\mu\nu}^{(n)} &= \sum_{klm}^{k+l+m=n} \langle \Phi|[h_{-\mu}^{(k)\dagger},[\tilde{H}^{(l)},h_{-\nu}^{(m)}]]_{+}|\Phi\rangle\label{eq:27}  \\
    S_{-\mu\nu}^{(n)} &= \sum_{kl}^{k+l=n} \langle \Phi|[h_{-\mu}^{(k)\dagger},h_{-\nu}^{(l)}]_{+}|\Phi\rangle\label{eq:28} \\
    T_{-p\nu}^{(n)} &= \sum_{kl}^{k+l=n} \langle \Phi|[\tilde{a}_{p}^{(k)},h_{-\nu}^{(l)}]_{+}|\Phi\rangle\label{eq:29} 
\end{align}
Here, the ionization operators $h_{-\mu}^{(k)\dagger}$ form the basis states $\ket{\Psi^{(k)}_{-\mu}} = h_{-\mu}^{(k)\dagger}\ket{\Phi}$ to represent electronic states of the $(N-1)$-electron system. The low-order components of these operators are given by $h_{-\mu}^{(0)\dagger} = a_{i}$ and ${h}_{-\mu}^{(1)\dagger} =a_a^{\dagger}a_ja_{i}$

Determining the matrix elements of $\textbf{M}_{\pm}$, $\textbf{S}_{\pm}$, and $\textbf{T}_{\pm}$ also requires solving for the $k$-th-order amplitudes of the excitation operators $T^{(k)}$. For the low-order EA- and IP-ADC(n) approximations (n $\le$ 3), equations for the transition energies and intensities depend on up to the $n$-th-order single- ($t_{i}^{a(n)}$) and $(n-1)$-th-order double-excitation ($t_{ij}^{ab(n-1)}$) amplitudes. At each order $k$, these amplitudes are computed by solving a system of projected amplitude equations:
\begin{align}
    \langle\Phi|a_{a}^{\dagger}a_{i}\tilde{H}^{(k)}|\Phi\rangle = 0\label{eq:25} \\
    \langle\Phi|a_{a}^{\dagger}a_{b}^{\dagger}a_{j}a_{i}\tilde{H}^{(k)}|\Phi\rangle = 0 \label{eq:26}
\end{align}

We present working equations for all matrix elements and amplitudes in \cref{eq:17,eq:18,eq:19,eq:27,eq:28,eq:29,eq:25,eq:26} of EA- and IP-ADC(n) (n $\le$ 3) in the Supporting Information. 

\section{Implementation}
\label{sec:implementation}
We implemented the EA- and IP-ADC(n) (n = 2, 3) methods in the development version of \textsc{Pyscf}.\cite{Sun:2018pe1340} Our program supports restricted and unrestricted reference Hartree-Fock orbitals and can be applied to closed- and open-shell molecules. We have verified that our IP-ADC(n) (n = 2, 3) program reproduces results from IP-ADC(n) implemented in \textsc{Q-Chem}.\cite{Dempwolff:2019p064108,qchem:44}  
Our EA- and IP-ADC(n) implementation features two algorithms for computing energies and density of states, which we briefly outline below. 

\subsection{Conventional algorithm}
\label{sec:implementation:conventional}
In the conventional algorithm, electron attachment or ionization energies are computed by solving the EA-/IP-ADC eigenvalue problem \eqref{eq:8} using iterative diagonalization techniques. In our implementation, we use a multi-root Davidson algorithm\cite{Davidson:1975p87,Liu:1978p49} that computes several lowest eigenvalues by starting with a set of guess (trial) eigenvectors $\mathbf{Y_\pm}$ and optimizing these vectors until convergence by forming the matrix-vector products $\boldsymbol{\sigma}_\pm \mathbf{ = M_\pm Y_\pm}$. To lower the computational cost of forming the $\boldsymbol{\sigma}_\pm$ vectors, matrix elements of the effective Hamiltonian (\cref{eq:17,eq:27}) with respect to the zeroth-order operators $h_{\pm\mu}^{(0)\dagger}$ (i.e., 1p-1p or 1h-1h block) are precomputed in the beginning of the Davidson procedure, stored in memory, and reused at every iteration. A similar approach has been recently taken by Dempwolff and co-workers\cite{Dempwolff:2019p064108} in the efficient implementation of IP-ADC(3). Once the Davidson diagonalization is complete, the eigenvalues and eigenvectors are used to compute the spectroscopic amplitudes $ \textbf{X}_{\pm}$ (\cref{eq:9}) and density of states $A(\omega)$ (\cref{eq:spec_function}) to obtain information about transition intensities and simulate photoelectron spectra. 


\subsection{Green's function algorithm}
\label{sec:implementation:dynamical}
In the second approach that we term the ``Green's function" algorithm (GF-ADC), the propagator and density of states are computed directly for a specific frequency $\omega$ bypassing the diagonalization of $\mathbf{M_\pm}$, in a way similar to that of Green's function coupled cluster methods.\cite{BhaskaranNair:2016p144101,McClain:2016p235139,Peng:2018p4335} Starting with \cref{eq:4}, expression for the ADC propagator can be written as:
\begin{align}
    \textbf{G}_\pm(\omega) = \textbf{Z}_\pm(\omega)\textbf{T}_\pm^{\dagger}\label{eq:g_xt}
    \end{align}
where the frequency-dependent $\mathbf{Z}_\pm(\omega)$ matrix is defined as
\begin{align}
    \mathbf{Z}_\pm(\omega)=\mathbf{T}_\pm(\omega\mathbf{S}_\pm-\mathbf{M}_\pm)^{-1}\label{eq:x_tsm}
\end{align}
Rearranging \cref{eq:x_tsm}, we obtain a system of linear equations
\begin{align}
   \textbf{T}_\pm^{\dag} - (\omega\textbf{S}_\pm-\textbf{M}_\pm)\textbf{Z}_\pm^{\dag}(\omega)=0\label{eq:t_s_mx}
\end{align}
that determines $\mathbf{Z}_\pm(\omega)$. Solving \cref{eq:t_s_mx} for $\mathbf{Z}_\pm(\omega)$ allows to compute the propagator $\textbf{G}_\pm(\omega)$ (\cref{eq:g_xt}) and density of states $A(\omega)$ (\cref{eq:spec_function}) at any frequency $\omega$ without explicitly diagonalizing $\mathbf{M_\pm}$. The GF-ADC algorithm is particularly well suited for computing densities of states at high excitation energies (e.g., ionization in core orbitals) or for simulating dense spectral regions with many electronic states in a narrow frequency region, which are expensive to compute using the conventional algorithm with traditional iterative techniques.

In our GF-ADC algorithm, we solve the linear equation using the conjugate gradient method,\cite{Shewchuk:1994p1} which requires forming the matrix-vector products $\boldsymbol{\sigma}_\pm (\omega) = \mathbf{ M_\pm Z_\pm^{\dag} (\omega)}$. For the frequency $\omega = \omega' + i\eta$ with a non-zero broadening parameter $\eta$, the elements of $\mathbf{Z}_\pm(\omega)$ and $\boldsymbol{\sigma}_\pm (\omega)$ are complex-valued. Aside from this, computing $\boldsymbol{\sigma}_\pm (\omega)$ is very similar to forming $\boldsymbol{\sigma}_\pm \mathbf{ = M_\pm Y_\pm}$ in the conventional algorithm, and both operations are implemented using the same general subroutine in our implementation. \cref{eq:t_s_mx} can be solved sequentially for a range of increasing or decreasing frequencies, reusing the solution $\mathbf{Z}_\pm(\omega)$ from the previous frequency as a guess for the next frequency computation, which significantly speeds up convergence of the conjugate gradient method. Alternatively, \cref{eq:t_s_mx} can be solved independently for each frequency, in a highly parallel fashion. Once computed, solutions $\mathbf{Z}_\pm(\omega)$ are used to evaluate the density of states $A(\omega)$ for a frequency region of interest. Interpolating  $A(\omega)$ between the frequency points, positions of the maxima are used to compute the corresponding electron attachment (or ionization) energies.

\section{Computational details}
\label{sec:comp_details}
We benchmarked the accuracy of EA- and IP-ADC(n) (n = 2, 3) for a variety of systems, including atoms (He -- Kr), small molecules (\cref{sec:results:ea_atoms_small_molecules}), five DNA/RNA nucleobases (\cref{sec:results:nucleobases}), zinc atom (\cref{sec:results:zinc}), and equally-spaced hydrogen chains (H$_n$ up to $n$ = 150, \cref{sec:results:hydrogen_chains}). 
Since the accuracy of IP-ADC(n) (n = 2, 3) has been thoroughly tested before,\cite{Trofimov:2005p144115,Dempwolff:2019p064108,Sokolov:2018p204113} for atoms, small molecules, and five nucleobases we only report the EA-ADC(n) (n = 2, 3) results. In addition to the strict EA-ADC(n) (n = 2, 3) approximations, we also present results for the extended EA-ADC(2) method (EA-ADC(2)-X) with partial treatment of third-order correlation effects, where the effective Hamiltonian $\tilde{H}$ in \cref{eq:17} is expanded up to the first order in perturbation theory with respect to the ${h}_{+\mu}^{(1)\dagger} =a_b^{\dagger}a_a^{\dagger}a_{i}$ operators. For the zinc atom, we use our GF-IP-ADC(n) implementation to compute core-level \mbox{$K$-,} $L$-, and $M$-shell ionization energies and compare these results with ionization energies from IP-ADC(n) with core-valence separation approximation [CVS-IP-ADC(n)]. For the hydrogen chains, we use EA- and IP-ADC(n) (n = 2, 3) to compute band gaps, as a difference between EA and IP. Throughout the manuscript, positive electron affinity implies exothermic electron attachment (i.e., EA = $E_{N}$ $-$ $E_{N+1}$), while a positive ionization energy corresponds to an endothermic process (IP = $E_{N-1}$ $-$ $E_{N}$).

The ADC results were compared to results from coupled cluster theory with single, double, and perturbative triple excitations [CCSD(T)] and equation-of-motion CCSD (EOM-CCSD).\cite{Crawford:2000p33,Shavitt:2009,Nooijen:1992p55,Nooijen:1993p15,Nooijen:1995p1681} For atoms and small molecules, all methods used the aug-cc-pVQZ basis set,\cite{Kendall:1992p6796} while the aug-cc-pVDZ basis was used for the nucleobases (\cref{sec:results:nucleobases}). Computations for the zinc atom were performed using the cc-pwCVTZ basis set\cite{Balabanov:2005p064107} combined with the spin-free one-electron variant of the X2C Hamiltonian.\cite{Liu:2010p1679,Saue:2011p3077,Peng:2012p1081} For the hydrogen chains, the STO-6G basis set was used.\cite{Hehre:1969p2657} Geometries of small molecules were optimized using CCSD(T)/aug-cc-pVQZ, while for molecules in \cref{sec:results:nucleobases} the equilibrium structures were computed using CCSD/aug-cc-pVDZ. The EOM-CCSD and CCSD(T) computations were performed using \textsc{Q-Chem},\cite{qchem:44} \textsc{Cfour},\cite{cfour} and \textsc{Psi4}.\cite{Parrish:2017p3185} The CVS-IP-ADC(n) (n = 2, 3) methods were implemented by restricting the operators $h_{-\mu}^{(k)\dagger}$ used to compute the $\mathbf{M_-}$ matrix (\cref{eq:27}) to $a_I$ and $a^\dag_a a_i a_I$, where $I$ denotes the index of the core orbital in the $K$-, $L$-, or $M$-shell.\cite{Angonoa:1987p6789,Schirmer:2001p10621,Thiel:2003p2088}

The GF-ADC computations for the zinc atom used the frequency interval ($\Delta\omega$) and broadening ($\eta$) of 0.005 \eh and 0.01 \eh, respectively. For computations using the conventional algorithm, intensities of EA and IP transitions were characterized by computing spectroscopic factors
\begin{align}
	\label{eq:spec_factors}
	P_{\pm\mu} = \sum_{p} |X_{\pm p\mu}|^2
\end{align}
where $X_{\pm p\mu}$ are elements of the spectroscopic amplitude matrix $\mathbf{X}_{\pm}$ defined in \cref{eq:9}. 

\section{Results}
\label{sec:results}

\begin{table*}[t!]
	\captionsetup{justification=raggedright,singlelinecheck=false,font=footnotesize}
	\caption{Vertical electron attachment energies ($\Omega$, eV) and spectroscopic factors ($P$) of closed-shell atoms computed using the aug-cc-pVQZ basis set. Also shown are mean absolute errors (\mae) and standard deviations (\std) of the results, relative to CCSD(T). }
	\label{tab:elec_affinity_atoms_closed_shell}
	\footnotesize
	\setstretch{1}
    \begin{tabular}{C{1.5cm}C{1.5cm}C{1.2cm}C{1.5cm}C{1.2cm}C{1.5cm}C{1.2cm}C{2cm}C{1.2cm}}
       \hline
        \hline
        \multicolumn{1}{c}{System} &\multicolumn{2}{c}{EA-ADC(2)} &\multicolumn{2}{c}{EA-ADC(2)-X} &\multicolumn{2}{c}{EA-ADC(3)} &\multicolumn{1}{c}{EA-EOM-CCSD} &\multicolumn{1}{c}{CCSD(T)} \\
        &$\Omega$ &$P$ &$\Omega$ &$P$ &$\Omega$ &$P$ &$\Omega$ &$\Omega$ \\
         \hline

He     		&$-$2.64  &1.00    &$-$2.62	&1.00   &$-$2.63        &1.00          &$-$2.63    &$-$2.63   \\     
Be      	&$-$0.25  &0.99    &$-$0.14 	&0.94   &$-$0.20        &0.95          &$-$0.27    &$-$0.27  \\      
Ne      	&$-$5.38  &0.99    &$-$5.31 	&0.99   &$-$5.30 	&0.99          &$-$5.34    &$-$5.29    \\    
Mg     		&$-$0.21  &0.98    &$-$0.13 	&0.95   &$-$0.17 	&0.95          &$-$0.23    &$-$0.22   \\     
Ar     		&$-$2.74  &0.98    &$-$2.67 	&0.98   &$-$2.78 	&0.98          &$-$2.77    &$-$2.82    \\    
Kr     		&$-$2.09  &0.98    &$-$2.02     &0.98   &$-$2.12 	&0.98          &$-$2.12    &$-$2.15   \\     
        
\mae            &0.046  &  &0.089 	&	&0.033 & 		&0.025    	&	 \\
\std        	&0.062  &  &0.072 	&     &0.030 &		&0.036   &		 \\	

        \hline
        \hline
    \end{tabular}
\end{table*}

\subsection{Electron affinities of atoms and small molecules}
\label{sec:results:ea_atoms_small_molecules}
We begin by benchmarking the accuracy of EA-ADC(2), EA-ADC(2)-X, and EA-ADC(3) for computing vertical electron attachment energies (EA's) of atoms (He -- Kr) and 16 small molecules. \cref{tab:elec_affinity_atoms_closed_shell,tab:elec_affinity_atoms_open_shell} present results of these methods for closed- and open-shell atoms, while EA's of closed- and open-shell molecules are reported in \cref{tab:elec_affinity_closed_shell_mol,tab:elec_affinity_open_shell_mol}, respectively.
We compare the ADC results with the electron affinities computed using EA-EOM-CCSD and reference data from CCSD(T). 
For all open-shell molecules, computed spin contamination of the unrestricted Hartree-Fock reference wavefunction does not exceed 0.05.

\begin{table*}[t!]
	\captionsetup{justification=raggedright,singlelinecheck=false,font=footnotesize}
	\caption{Vertical electron attachment energies ($\Omega$, eV) and spectroscopic factors ($P$) of open-shell atoms computed using the aug-cc-pVQZ basis set. Experimental values are obtained from Ref.\@ \citenum{Rumble:2019crc}. Also shown are mean absolute errors (\mae) and standard deviations (\std) of the results, relative to CCSD(T), and spin contamination of the reference unrestricted Hartree-Fock wavefunction  ($\langle \hat{S}^{2}\rangle-S(S+1)$).}
	\label{tab:elec_affinity_atoms_open_shell}
	\footnotesize
\begin{tabular}{C{1.0cm}C{1.8cm}C{1.0cm}C{1.0cm}C{1.0cm}C{1.0cm}C{1.0cm}C{1.0cm}C{0.8cm}C{2.5cm}C{1.4cm}C{0.8cm}C{0.8cm}}
       \hline
        \hline
        \multicolumn{1}{c}{System} &\multicolumn{1}{c}{State}
        &\multicolumn{1}{c}{Spin cont.} 
        &\multicolumn{2}{c}{EA-ADC(2)} &\multicolumn{2}{c}{EA-ADC(2)-X} &\multicolumn{2}{c}{EA-ADC(3)} &\multicolumn{1}{c}{EA-EOM-CCSD} &\multicolumn{1}{c}{CCSD(T)} &\multicolumn{1}{c}{Exp.}\\
        & & &$\Omega$ &$P$ &$\Omega$ &$P$ &$\Omega$ &$P$ &$\Omega$ &$\Omega$ &$\Omega$ \\
         \hline

Li    &${}^{2}S\rightarrow {}^{1}S$ &0.000 	&0.34  	  &0.94    &0.62 	&0.78       &0.62 	&0.78         &0.61   	  &0.62      &0.62 \\           
B     &${}^{2}P\rightarrow {}^{3}P$ &0.011 		&0.07  	  &0.96    &0.51 	&0.85       &0.32 	&0.85         &0.14   	  &0.25      &0.28 \\           
C     &${}^{3}P\rightarrow {}^{4}S$ &0.010 		&0.98  	  &0.95    &1.61 	&0.85       &1.24 	&0.86         &1.10   	  &1.25      &1.26 \\           
N &${}^{4}S\rightarrow {}^{3}P$ &0.008  &$-$0.68 &0.96 &0.05 &0.85 &$-$0.38 &0.87 &$-$0.34 &$-$0.23 &\\
O     &${}^{3}P\rightarrow {}^{2}P$ &0.009		&0.90  	  &0.94    &1.83 	&0.84       &1.06 	&0.86         &1.22   	  &1.40      &1.46 \\           
F      &${}^{2}P\rightarrow {}^{1}S$ &0.004		&3.09  	  &0.93    &4.00 	&0.84       &2.90 	&0.87         &3.23   	  &3.39      &3.40 \\           
Na    &${}^{2}S\rightarrow {}^{1}S$ &0.000	&0.33  	  &0.95    &0.56 	&0.79       &0.56 	&0.79         &0.54   	  &0.55      &0.55 \\           
Al     &${}^{2}P\rightarrow {}^{3}P$ &0.020 	&0.33  	  &0.96    &0.58 	&0.87       &0.47 	&0.87         &0.34  	  &0.43      &0.43 \\           
Si    &${}^{3}P\rightarrow {}^{4}S$ &0.015 		&1.32  	  &0.95    &1.63 	&0.88       &1.42 	&0.88         &1.29   	  &1.40      &1.39 \\           
P      &${}^{4}S\rightarrow {}^{3}P$ &0.001		&0.51     &0.94    &0.97 	&0.87       &0.65 	&0.87         &0.60   	  &0.69      &0.75 \\           
S      &${}^{3}P\rightarrow {}^{2}P$ &0.013		&1.90  	  &0.94    &2.40 	&0.87       &1.94 	&0.88         &1.92   	  &2.04      &2.08 \\           
Cl     &${}^{2}P\rightarrow {}^{1}S$ &0.010		&3.59  	  &0.93    &4.07 	&0.88       &3.46 	&0.89         &3.49   	  &3.61      &3.61 \\           
Ga     &${}^{2}P\rightarrow {}^{3}P$ &0.013		&0.28     &0.96    &0.51 	&0.88       &0.47 	&0.88         &0.27   	  &0.35      &0.43 \\           
Ge     &${}^{3}P\rightarrow {}^{4}S$ &0.011		&1.31  	  &0.95    &1.59 	&0.89       &1.49 	&0.89         &1.29   	  &1.38      &1.23 \\           
As     &${}^{4}S\rightarrow {}^{3}P$ &0.001		&0.55  	  &0.95    &0.95 	&0.87       &0.77 	&0.88         &0.64   	  &0.73      &0.80 \\           
Se     &${}^{3}P\rightarrow {}^{2}P$ &0.012		&1.89  	  &0.94    &2.31 	&0.88       &2.04 	&0.89         &1.92   	  &2.04      &2.02 \\           
Br     	&${}^{2}P\rightarrow {}^{1}S$ &0.010	&3.44  	  &0.94    &3.83 	&0.88       &3.46 	&0.90         &3.37   	  &3.50      &3.36 \\ 
\mae       &  &     &0.189 &  &0.274  	&  &0.101 &	&0.102 &	& \\
\std       & & 	   &0.137  &  &0.152 	&  &0.158 &	&0.045   &		& \\	
        \hline
        \hline
    \end{tabular}
\end{table*}

For all atoms (\cref{tab:elec_affinity_atoms_closed_shell,tab:elec_affinity_atoms_open_shell}), EA-ADC(3) and EA-EOM-CCSD produce results in a very good agreement with CCSD(T). In the case of closed-shell atoms, both EA-ADC(3) and EA-EOM-CCSD show very similar accuracy with mean absolute errors (\mae) and standard deviations (\std) of $\sim$ 0.03 eV, relative to CCSD(T).  For atoms with unpaired electrons, all EA-ADC and EA-EOM-CCSD methods produce higher \mae errors in vertical EA's compared to the closed-shell ones. The best agreement with CCSD(T) is once again shown by EA-ADC(3) and EA-EOM-CCSD with a similar \mae of $\sim$ 0.1 eV. The spectroscopic factors of open-shell atoms ($P$, \cref{tab:elec_affinity_atoms_closed_shell,tab:elec_affinity_atoms_open_shell}) show significant dependence on the order of the ADC approximation, with the EA-ADC(2) values of $\sim$ 0.95 and the EA-ADC(3) results of $\sim$ 0.85, indicating importance of electron correlation effects.

\begin{table*}[t!]
	\captionsetup{justification=raggedright,singlelinecheck=false,font=footnotesize}
    \caption{Vertical electron attachment energies ($\Omega$, eV) and spectroscopic factors ($P$) of small closed-shell molecules computed using the aug-cc-pVQZ basis set. Also shown are mean absolute errors (\mae) and standard deviations (\std) of the results, relative to CCSD(T). }
	\label{tab:elec_affinity_closed_shell_mol}
	\setstretch{1}
	\footnotesize
\begin{tabular}{C{2cm}C{1.5cm}C{1.2cm}C{1.5cm}C{1.2cm}C{1.5cm}C{1.2cm}C{2.5cm}C{1.5cm}}
       \hline
        \hline
        \multicolumn{1}{c}{System} &\multicolumn{2}{c}{EA-ADC(2)} &\multicolumn{2}{c}{EA-ADC(2)-X} &\multicolumn{2}{c}{EA-ADC(3)} &\multicolumn{1}{c}{EA-EOM-CCSD} &\multicolumn{1}{c}{CCSD(T)} \\
        &$\Omega$ &$P$ &$\Omega$ &$P$ &$\Omega$ &$P$ &$\Omega$ &$\Omega$ \\
         \hline
        
LiH          	&0.28    &0.99     &0.32	&0.99             &0.30        &0.99    &0.29      &0.29 \\
SiH$_{2}$     	&1.16    &0.95	   &1.40 	&0.90             &1.09 	  &0.90    &1.00      &1.03  \\
BeO            	&2.30    &0.97	   &2.37 	&0.96             &2.05 	  &0.96    &2.15  	   &2.13   \\
CO            	&$-$1.41 &0.98     &$-$1.29     &0.95 	          &$-$1.41     &0.99	   &$-$1.46   &$-$1.44   \\
H$_2$O         	&$-$0.57 &0.99     &$-$0.54     &0.99 	          &$-$0.57     &0.99    &$-$0.57   &$-$0.56   \\
SO$_2$	        &1.09    &0.91     &1.27	&0.88             &0.92	  &0.90    &0.85	   &0.84    \\
HF            	&$-$0.64 &1.00     &$-$0.61     &0.99	          &$-$0.63     &0.99	   &$-$0.63   &$-$0.62   \\
N$_2$         	&$-$1.81 &0.99	   &$-$1.79     &0.99	          &$-$1.84     &0.99	   &$-$1.83   &$-$1.83 \\
P$_2$         	&0.74    &0.90	   &0.89 	&0.87             &0.56 	  &0.88    &0.44      &0.45  \\
F$_2$          	&0.47    &0.91     &0.91 	&0.86             &0.34 	  &0.89    &0.13      &0.34  \\
H$_2$CO         &$-$0.51 &0.99     &$-$0.48 &0.99	          &$-$0.56     &0.99	   &$-$0.53   &$-$0.55\\ 

\mae            &0.100	 &      &0.214	 &  &0.038	 & &0.031      &  \\
\std            &0.110	 &      &0.204	 &  &0.054	 & &0.061   &    \\
        \hline
        \hline
    \end{tabular}
\end{table*}

Accuracy of the EA-ADC methods for closed-shell molecules (\cref{tab:elec_affinity_closed_shell_mol}) is consistent with that observed for atoms with no unpaired electrons, where EA-ADC(3) and EA-EOM-CCSD yield the best results showing similar \mae and \std in the range of 0.03 - 0.06 eV. However, performance of the EA-ADC methods for the five open-shell radicals (\cref{tab:elec_affinity_open_shell_mol}) is in a stark contrast to that for the open-shell atoms. In this case, out of the four methods the best agreement with CCSD(T) is given by EA-ADC(2) with \mae and \std of 0.14 and 0.11 eV, respectively. The EA-EOM-CCSD method shows a significantly larger \mae = 0.25 eV, followed by EA-ADC(3) with \mae = 0.37 eV. This deviation from previous trends may be explained by the combination of the poor performance of perturbation theory for open-shell systems along with a fortuitous error cancellation in the computed EA-ADC(2) values. 

\begin{table*}[t!]
	\captionsetup{justification=raggedright,singlelinecheck=false,font=footnotesize}
    \caption{Vertical electron attachment energies ($\Omega$, eV) and spectroscopic factors ($P$) of small open-shell systems computed using the aug-cc-pVQZ basis set. Also shown are mean absolute errors (\mae) and standard deviations (\std) of the results, relative to CCSD(T), and spin contamination of the reference unrestricted Hartree-Fock wavefunction  ($\langle \hat{S}^{2}\rangle-S(S+1)$). }
	\label{tab:elec_affinity_open_shell_mol}
	\footnotesize
\begin{tabular}{C{1cm}C{1.8cm}C{1.2cm}C{1cm}C{1cm}C{1cm}C{1cm}C{1cm}C{1cm}C{2.5cm}C{1.4cm}}
       \hline
        \hline
        \multicolumn{1}{c}{System} &\multicolumn{1}{c}{State}
        &\multicolumn{1}{c}{Spin cont.}
        &\multicolumn{2}{c}{EA-ADC(2)} &\multicolumn{2}{c}{EA-ADC(2)-X} &\multicolumn{2}{c}{EA-ADC(3)} &\multicolumn{1}{c}{EA-EOM-CCSD} &\multicolumn{1}{c}{CCSD(T)} \\
        &  & &$\Omega$ &$P$ &$\Omega$ &$P$ &$\Omega$ &$P$ &$\Omega$ &$\Omega$ \\
         \hline
OH       &${}^{2}\Pi \rightarrow {}^{1}\Sigma^{+}$       &0.007     	&1.70    &0.92   &2.39	&0.83    &1.23        &0.87       &1.61     &1.80 \\
NH       &${}^{3}\Sigma^{-}\rightarrow {}^{2}\Pi  $       &0.017	&0.01    &0.94	 &0.67 	&0.85    &$-$0.02     &0.87       &0.14     &0.29  \\
O$_2$    &${}^{3}\Sigma^{-}_{g}\rightarrow {}^{2}\Pi_{g}$ &0.049        &$-$0.34 &0.92	 &0.15 	&0.86    &$-$0.35     &0.89       &$-$0.34  &$-$0.13   \\
NH$_2$   &${}^{2}B_{1}\rightarrow {}^{1}A_{1}$            &0.010     	&0.70    &0.93   &1.22  &0.84    &0.24        &0.88       &0.56     &0.71   \\
CH$_3$	 &${}^{2}A''_{2}\rightarrow {}^{1}A'_{1}$         &0.012        &$-$0.21  &0.95  &0.23	&0.84    &$-$0.37     &0.86       &$-$0.66  &$-$0.09\\
\mae     & &       &0.140	 &   &0.416	 & &0.369	 &     &0.252      &  \\
\std     &  &     &0.105	 &   &0.131	 & &0.145	 &     &0.179   &    \\
        \hline
        \hline
    \end{tabular}
\end{table*}

\begin{figure*}[t!]
     \subfigure[]{\includegraphics[width=2in]{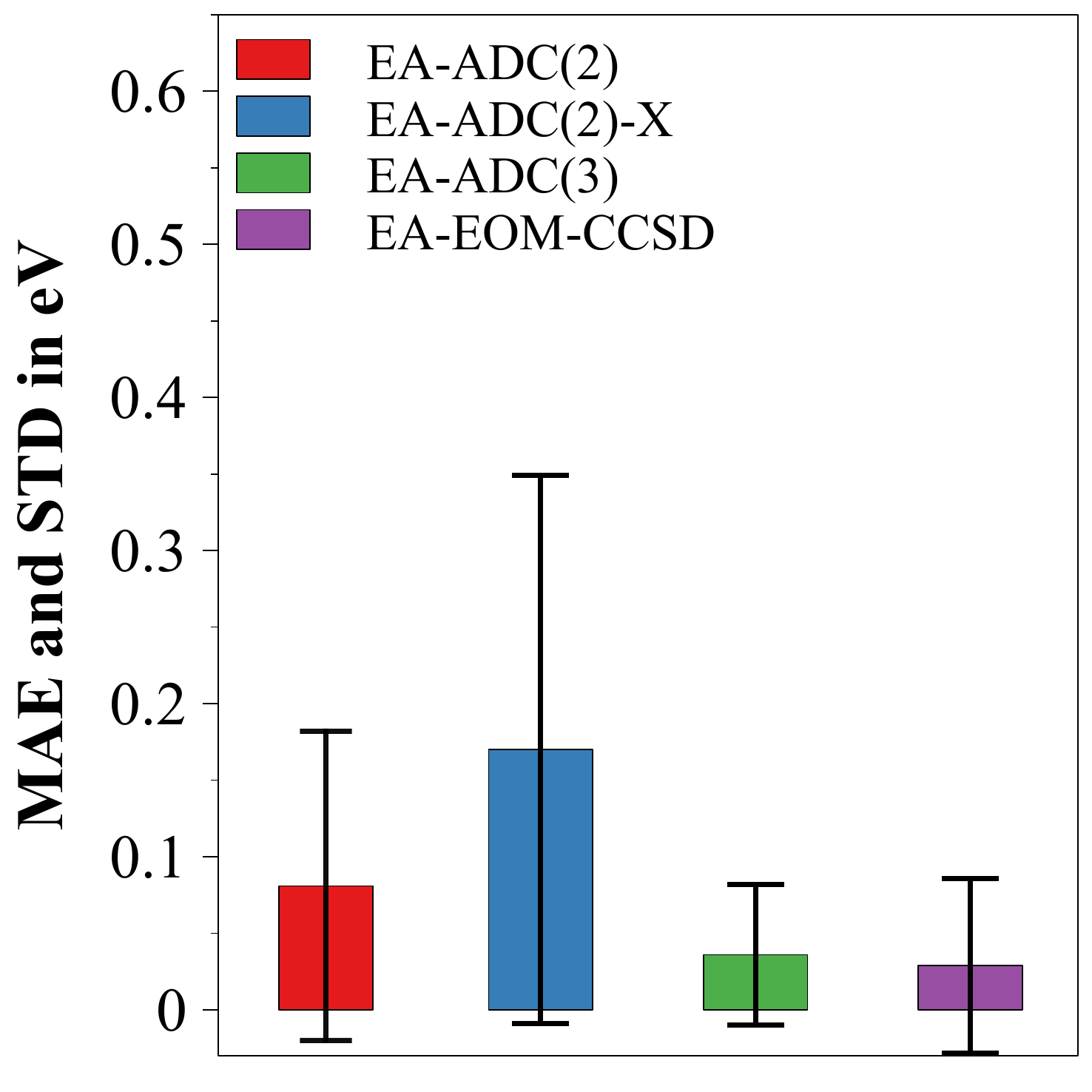}\label{fig:atoms_and_mol_stat_closed}}  \qquad
    \subfigure[]{\includegraphics[width=2in]{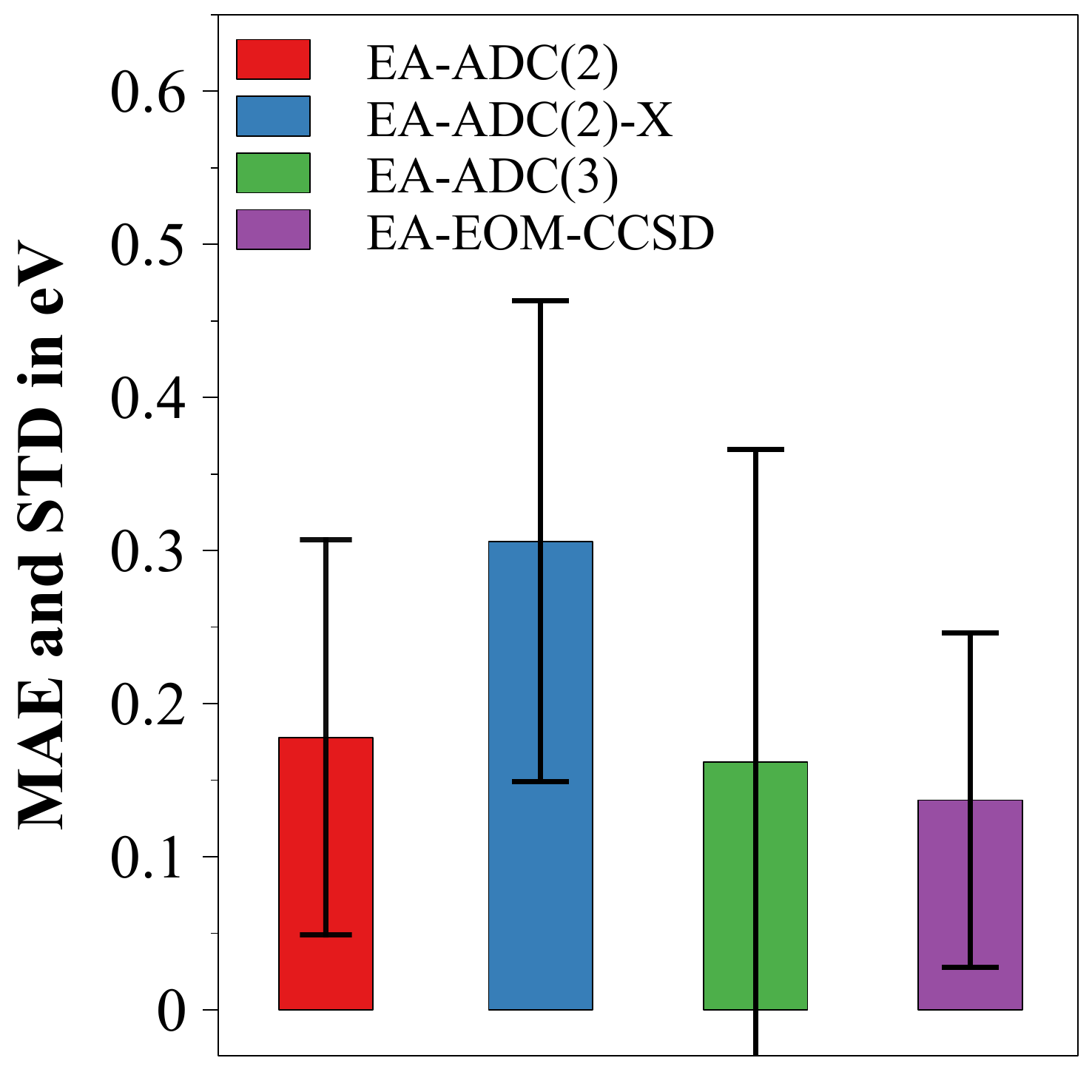}\label{fig:atoms_and_mol_stat_open}} 
	\captionsetup{justification=raggedright,singlelinecheck=false,font=footnotesize}
	\caption{Mean absolute errors (MAE, eV) and standard deviations from the mean signed error (STD, eV) for vertical attachment energies of (a) closed-shell and (b) open-shell atoms and molecules computed using four methods, relative to CCSD(T) (aug-cc-pVQZ basis set). The MAE value is represented as a height of each colored box, while the STD value is depicted as a radius of the black vertical bar. See \cref{tab:elec_affinity_atoms_closed_shell,tab:elec_affinity_atoms_open_shell,tab:elec_affinity_closed_shell_mol,tab:elec_affinity_open_shell_mol} for data on individual molecules.}
	\label{fig:atoms_and_mol_stat}
\end{figure*}          
			
\cref{fig:atoms_and_mol_stat_closed,fig:atoms_and_mol_stat_open} demonstrate overall performance of the EA-ADC methods for atoms and molecules combined together. On average, performance of EA-ADC(3) and EA-EOM-CCSD is similar for closed- and open-shell systems, although the former method is less reliable for systems with unpaired electrons, which is evidenced by its larger standard deviation. Interestingly, we find that for most systems the extended EA-ADC(2)-X scheme produces larger errors compared to EA-ADC(2), despite incorporating description of the higher-order electron correlation effects. This observation is consistent with the performance of IP-ADC(2)-X for ionization energies,\cite{Trofimov:2005p144115} indicating unbalanced nature of this approximation.

\begin{table*}[t!]
	\captionsetup{justification=raggedright,singlelinecheck=false, font=footnotesize}
    \caption{Errors in vertical electron attachment energies (EA's, eV), mean absolute errors (\mae) and standard deviations (\std) of DNA/RNA nucleobases computed using four methods with the aug-cc-pVDZ basis set, relative to EA's from CCSD(T) shown in the rightmost column.}
	\label{tab:elec_affinity_poly}
	\setstretch{1}
	\footnotesize
\begin{tabular}{C{2cm}C{2.5cm}C{2.5cm}C{2.5cm}C{3.0cm}C{2cm}}
       \hline
        \hline
        \multicolumn{1}{c}{System} &\multicolumn{1}{c}{EA-ADC(2)} &\multicolumn{1}{c}{EA-ADC(2)-X} &\multicolumn{1}{c}{EA-ADC(3)} &\multicolumn{1}{c}{EA-EOM-CCSD} &\multicolumn{1}{c}{CCSD(T)} \\
        
         \hline
Uracil          &0.09	&0.15	&$-$0.05 &0.02     &$-$0.25\\
Cytosine        &0.05	&0.11	&$-$0.05 &0.00     &$-$0.27\\
Thymine         &0.08	&0.14	&$-$0.06 &0.01     &$-$0.29\\
Adenine         &0.02	&0.08	&$-$0.04 &$-$0.02   &$-$0.42  \\
Guanine        &0.02	&0.09	&$-$0.04 &$-$0.03    &$-$0.17 \\
\mae            &0.05	 &0.11  &0.05           &0.02  &  \\
\std            &0.03	 &0.03  &0.01   	 &0.02     &  \\
        \hline
        \hline
    \end{tabular}
\end{table*}

\subsection{Electron attachment in DNA/RNA nucleobases}
\label{sec:results:nucleobases}
In this section, we extend our benchmark of the EA-ADC methods to five nucleic acid bases present in DNA and RNA: uracil (U), thymine (T), cytosine  (C), adenine (A), and guanine (G). Accurate simulation of EA's of these systems is crucial for understanding electron attachment in DNA and RNA that can induce biochemical processes such as charge transfer along the DNA/RNA strand and alteration of genetic information.\cite{Steenken:1989p503,Yan:1992p1983,Colson:1995p3867,Wintjens:2000p393,Huels:1998p1309} Experimentally, electron attachment of the isolated nucleobases has been studied by electron transmission spectroscopy (ETS)\cite{Jordan:1987p557,Aflatooni:1998p6205,Periquet:2000p141} for all systems except guanine, whose naturally stable keto isomer becomes unstable in the gas phase. The ETS measurements yield negative EA's for the four remaining nucleobases that become increasingly negative as U $>$ T $>$ C $>$ A,\cite{Aflatooni:1998p6205} indicating that for all of these systems electron attachment is an endothermic process in the gas phase. Theoretically, EA's of the DNA/RNA nucleobases have been a subject of a number of investigations.\cite{Roca:2008p095104,Roca:2006p084302,Li:2002p1596,Wesolowski:2001p4023,Sevilla:1995p1060,Richardson:2002p10163,Chen:2000p7835,Dutta:2015p753,Dedikova:2009p107} We specifically mention the CCSD(T) study of (U, T, C) by Roca-Sanju\'an et al.\cite{Roca:2008p095104} and the EA-EOM-CCSD study of all five nucleobases by Dutta et al.\cite{Dutta:2015p753} Although both studies were carried out using the aug-cc-pVDZ basis set, vertical EA's of the (U, T, C) nucleobases computed using CCSD(T) and EA-EOM-CCSD are significantly different: ($-$0.64, $-$0.65, $-$0.79) and  ($-$0.23, $-$0.28, $-$0.27) eV, respectively. Another CCSD(T) study by Urban and co-workers\cite{Dedikova:2009p107} reports adiabatic EA for uracil of $-$0.19 eV, in a close agreement with the EA-EOM-CCSD vertical EA of $-$0.23 eV computed by Dutta et al.\cite{Dutta:2015p753}

To resolve the disagreement between the CCSD(T) and EA-EOM-CCSD results in the literature, we recomputed vertical EA's of the five nucleobases using these two methods with the aug-cc-pVDZ basis set (\cref{tab:elec_affinity_poly}). For all systems, our CCSD(T) and EA-EOM-CCSD vertical EA's are in a very close agreement between each other (with the largest difference of 0.03 eV) and with the EA-EOM-CCSD results reported by Dutta et al. For all $C_{s}$-symmetric molecules (U, C, T, G), we find that the ground electronic state of the anion is $^2A'$. 
When computing the CCSD(T) vertical electron affinity with respect to the lowest-energy $^2A''$ state of the anion, we obtain EA values that are close to those reported by Roca-Sanju\'an et al.,\cite{Roca:2008p095104} which suggests a possible source for the disagreement with the results of this study. 

\begin{figure}[t]
\centering
\includegraphics[width=3in]{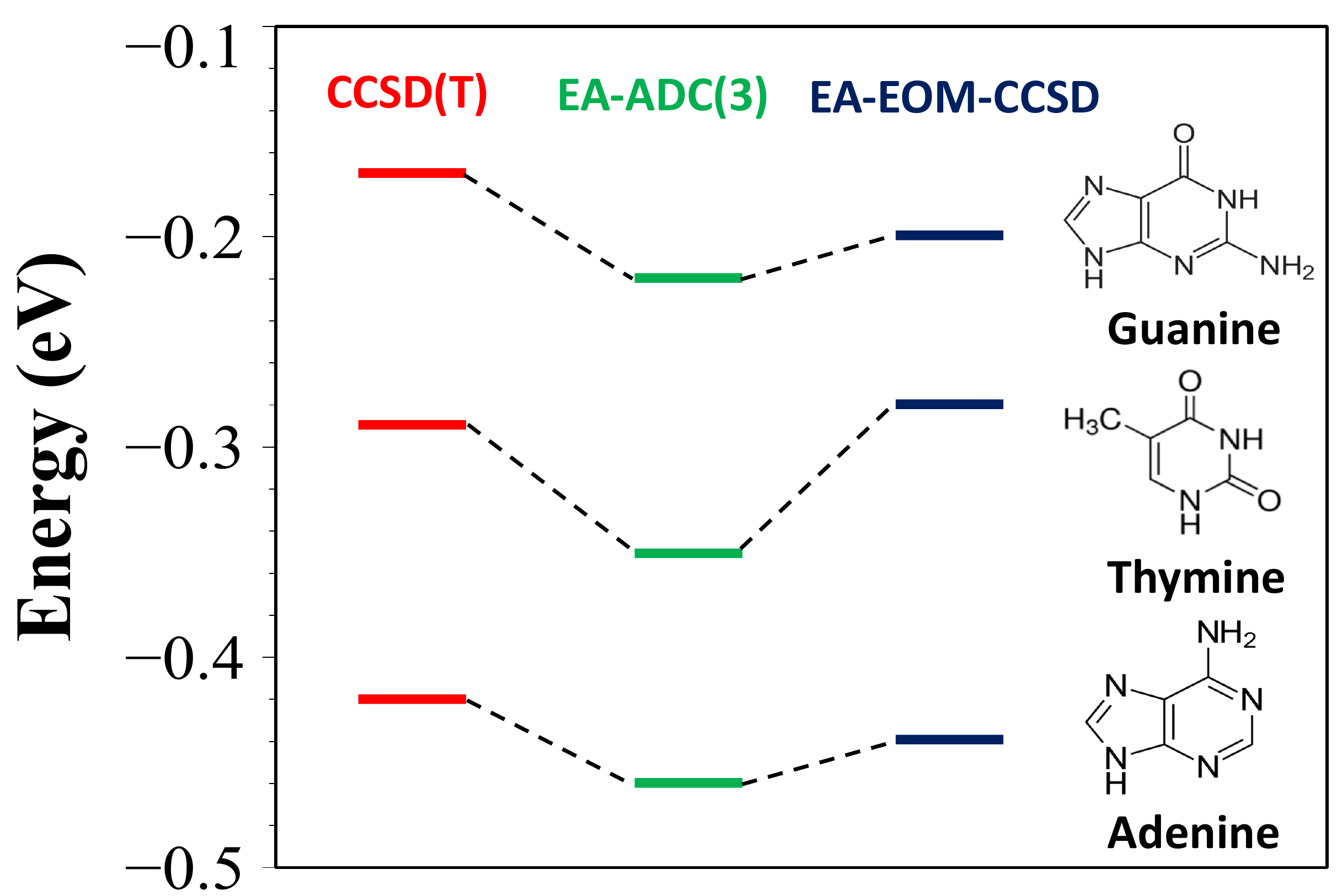}
\captionsetup{justification=raggedright,singlelinecheck=false,font=footnotesize}
  \caption{Vertical electron attachment energies (in eV) of DNA/RNA nucleobases computed using three different methods with the aug-cc-pVDZ basis set.}
   \label{fig:bases}
\end{figure}

\cref{tab:elec_affinity_poly} presents errors in the vertical EA's computed using EA-ADC and EA-EOM-CCSD, relative to CCSD(T). Consistent with our previous results for atoms and small molecules (\cref{sec:results:ea_atoms_small_molecules}), the best agreement with CCSD(T) is shown by EA-EOM-CCSD with \mae and \std of 0.02 eV. The EA-ADC(3) method shows a somewhat larger \mae = 0.05 eV, but smaller \std = 0.01 eV, systematically underestimating the CCSD(T) values. The EA-ADC(2) approximation shows similar accuracy, overestimating the reference energies with small \mae = 0.05 eV and \std = 0.03 eV. The EA's computed using CCSD(T), EA-EOM-CCSD, and EA-ADC(3) become increasingly negative in the order G $>$ U $>$ C $>$ T $>$ A, which agrees with the experimental trend for all nucleobases but thymine and cytosine, where T $>$ C.\cite{Aflatooni:1998p6205} Given the fact that EA's of these two nucleobases are very close, their relative order may be affected by incompleteness of the basis set and level of dynamic correlation treatment. When comparing energy differences between EA's ($\Delta$EA) of different nucleobases, the EA-ADC(3) method shows better agreement with CCSD(T) than EA-EOM-CCSD. As illustrated in \cref{fig:bases}, EA-EOM-CCSD underestimates $\Delta$EA for G and T by 0.04 eV and overestimates $\Delta$EA between T and A by 0.03 eV, while the EA-ADC(3) shows errors of 0.01 and 0.02 eV for the respective transitions.

\subsection{Core ionization energies of the zinc atom}
\label{sec:results:zinc}

\begin{table*}[t!]
	\captionsetup{justification=raggedright,singlelinecheck=false, font=footnotesize}
    \caption{Core-level ionization energies (eV) of the zinc atom computed using IP-ADC with core-valence separation approximation (CVS-IP-ADC) and unapproximated Green's function IP-ADC (GF-IP-ADC) with the cc-pwCVTZ basis set. For each method, results of the relativistic (X2C) and nonrelativistic (NR) computations are shown. Experimental values are from Ref.\@ \citenum{Ley:1973p2392}. }
	\label{tab:zinc_IP}
	\setstretch{1.0}
	\footnotesize
	\begin{threeparttable}
\begin{tabular}{C{0.8cm}C{0.8cm}C{1.2cm}C{1.5cm}C{1.2cm}C{1.5cm}C{1.2cm}C{1.5cm}C{1.2cm}C{1.5cm}C{1.2cm}}
       \hline
        \hline
        \multicolumn{1}{c}{Shell} &\multicolumn{1}{c}{Orbital} &\multicolumn{2}{c}{GF-IP-ADC(2)} &\multicolumn{2}{c}{CVS-IP-ADC(2)} &\multicolumn{2}{c}{GF-IP-ADC(3)} &\multicolumn{2}{c}{CVS-IP-ADC(3)} &\multicolumn{1}{c}{Experiment}\\
        & &NR &X2C &NR &X2C &NR &X2C &NR  &X2C\\
        \hline
$K$       &$1s$   &9555.44    &9648.48	&9555.47	&9648.51	&9584.80	&9678.12	&9584.77	&9678.12 		&\\
$L$       &$2s$  &1164.18     &1183.97	&1170.50	&1193.03	&1185.87	&1206.96  &1195.04		&1218.04		&1196.16 \\
$L$        &$2p$   &1021.27    &1027.64	&1021.62	&1028.67	&1044.89	&1052.24  &1046.30		&1053.76	&1033.52\tnote{a} \\
$M$       &$3s$  &138.40     &141.17 &142.56	 	&146.02	    &148.63	&152.11	 &151.70   	&155.59	   	&139.88\\
$M$       &$3p$   &91.48    &92.33	  &94.86  	&95.76	 &99.92   	&101.09	    &102.51	&103.73	 	&90.01\tnote{a}\\
        \hline
        \hline
    \end{tabular}
    \begin{tablenotes}
    \item[a] Experimental values for ionization in the $2p$ and $3p$ orbitals are obtained by averaging ionization energies for states with the total angular momentum quantum number $J$ = 1/2 and 3/2.
    \end{tablenotes}
    \end{threeparttable}
\end{table*}

We now demonstrate capabilities of our Green's function IP-ADC implementation (GF-IP-ADC) by computing core-level ionization energies of the zinc atom as a prototypical first-row transition metal with closed-shell neutral ground state. The IP-ADC method has been previously used to compute core ionization energies\cite{Angonoa:1987p6789,Schirmer:2001p10621,Thiel:2003p2088} in combination with core-valence separation (CVS) approximation\cite{Cederbaum:1980p481,Cederbaum:1980p206} (CVS-IP-ADC), which neglects coupling between the core and valence ionized states in the effective Hamiltonian matrix $\mathbf{M_-}$ due to their large energetic separation. In this study, we use GF-IP-ADC to compute the $K$-, $L$-, and $M$-shell ionization energies of the zinc atom without introducing the CVS approximation. We compare our results to ionization energies computed using the CVS-IP-ADC methods and assess accuracy of the CVS approximation for ionization in each electronic shell. 

\cref{tab:zinc_IP} compares results of GF-IP-ADC and CVS-IP-ADC for the $K$-, $L$-, and $M$-shell ionization energies along with experimental data for the metallic zinc.\cite{Ley:1973p2392} All IP-ADC computations employed the cc-pwCVTZ basis set and the spin-free exact two-component (X2C) Hamiltonian to account for scalar relativistic effects. To assess effect of the relativistic effects on the CVS errors, \cref{tab:zinc_IP} also presents results of the IP-ADC methods without using the X2C Hamiltonian. The $K$-shell ionization energies computed using GF-IP-ADC and CVS-IP-ADC are in a very close agreement, suggesting that the CVS approximation is very accurate for ionization in the $1s$ orbital. However, significant errors of the CVS approximation are observed for ionization in the $L$ and $M$ shells, with or without accounting for relativistic effects, ranging from $\sim$ 0.35 eV for the $2p$ orbital up to $\sim$ 11 eV for the $2s$ orbital. For all of these transitions, the CVS-IP-ADC results overestimate the unapproximated GF-IP-ADC energies. The magnitude of the CVS errors does not correlate with excitation energy, the orbital angular momentum, or the order of perturbation theory in relativistic and non-relativistic computations, indicating an uncontrolled nature of the CVS approximation. Incorporating relativistic effects contributes to increasing the CVS errors by at most $\sim$ 30 $\%$. Avoiding the CVS approximation in GF-IP-ADC(3) significantly reduces errors in ionization energies relative to the experiment. For example, for the $2s$ orbital the relativistic GF-IP-ADC(3) and CVS-IP-ADC(3) errors are 11.8 and 22.9 eV, respectively. When considering relative errors, the CVS approximation has the largest effect on the $M$-shell ionization energies, overestimating their values by up to $\sim$ 4 \%. 

\begin{figure*}[t!]
     \subfigure[]{\includegraphics[width=2.5in]{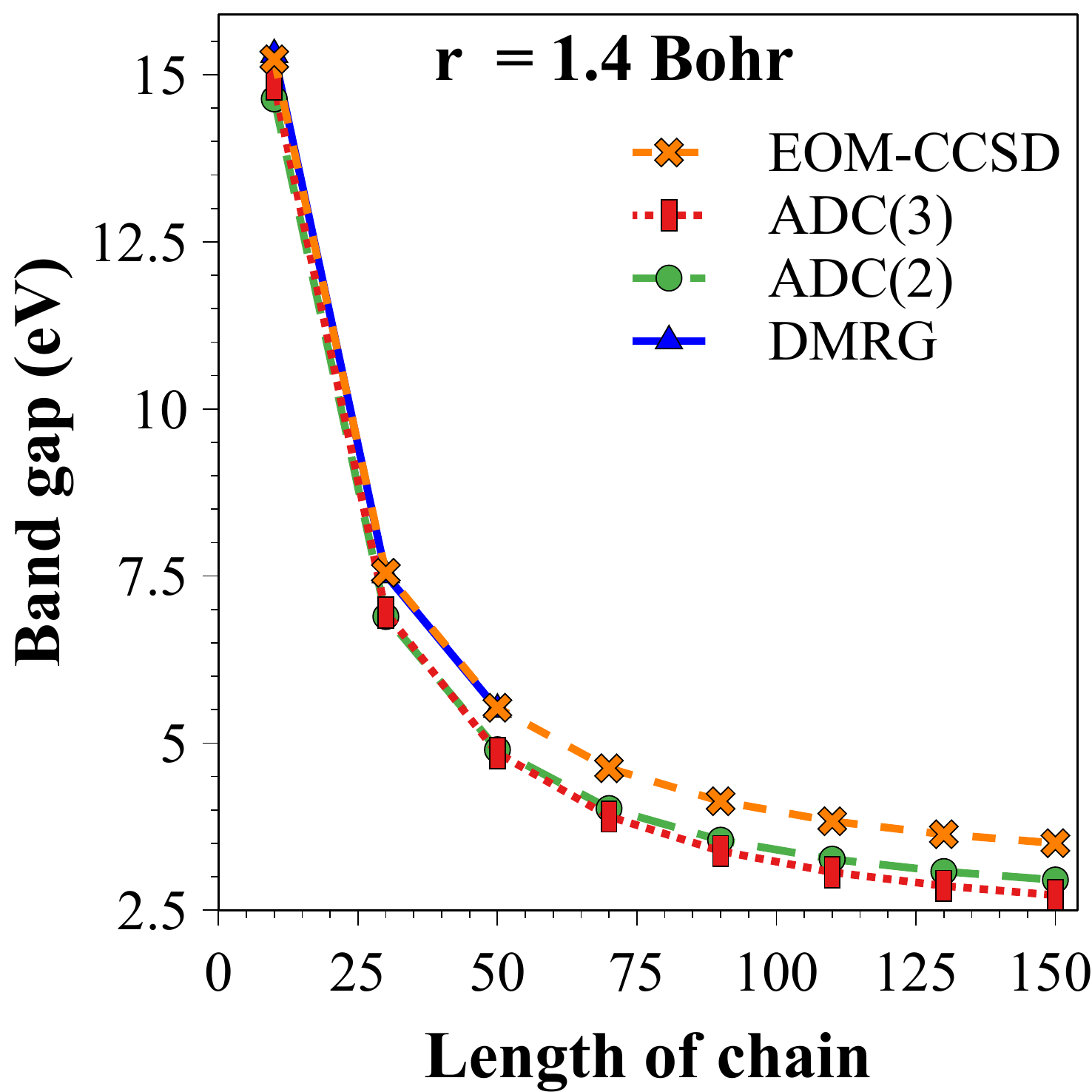}\label{fig:band_gap_1.4}}  \qquad
    \subfigure[]{\includegraphics[width=2.5in]{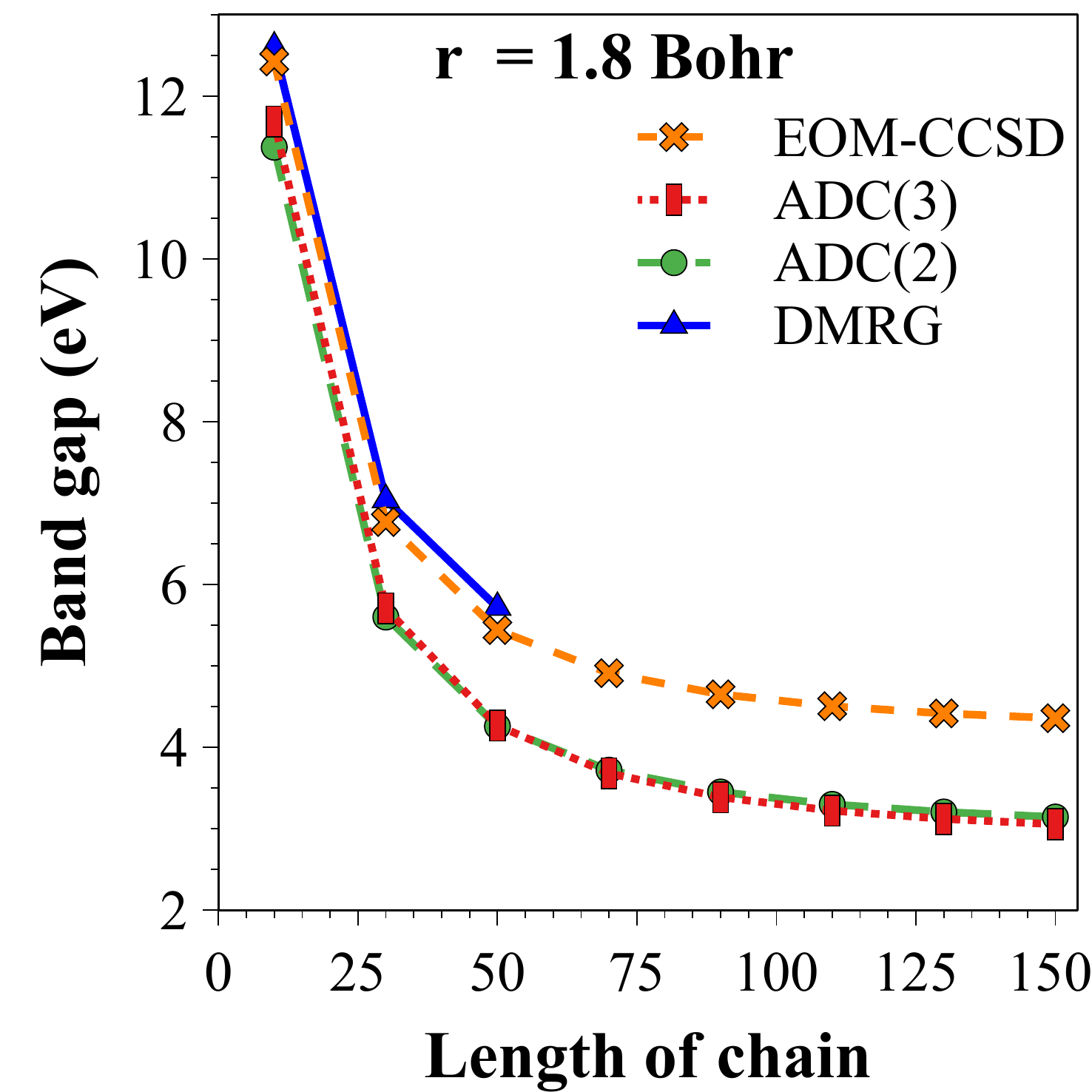}\label{fig:band_gap_1.8}} 
	\captionsetup{justification=raggedright,singlelinecheck=false,font=footnotesize}
	\caption{Band gaps (in eV) of the equally-spaced hydrogen chains (H$_n$) as a function of the chain length ($n$ = 10 - 150) and H--H bond distance ($r$ = 1.4 and 1.8 $a_0$) computed using four methods with the STO-6G basis set. The DMRG results are from Ref.\@ \citenum{Ronca:2017p5560}. }
	\label{fig:band_gap}
\end{figure*}

\subsection{Band gaps of hydrogen chains}
\label{sec:results:hydrogen_chains}
Finally, we combine our EA- and IP-ADC implementations to study equally-spaced hydrogen chains (H$_n$, $n$ = 10 - 150), which are challenging one-dimensional models for understanding strong electron correlation in molecules and materials.\cite{Sinitskiy:2010p014104,Stella:2011p245117,Lin:2011p096402,Tsuchimochi:2009p121102,Rusakov:2016p054106,Motta:2017p031059,Rusakov:2019p229} An important property of hydrogen chains is their band gap that can be calculated as the difference between IP and EA. In the thermodynamic limit, hydrogen chains are believed to exist in two phases: (i) a metallic phase with a zero band gap corresponding to short H--H distances ($r$) and (ii) an insulator phase with a non-zero band gap at longer distances. Recently, Ronca et al.\cite{Ronca:2017p5560} computed band gaps for H$_n$ ($n$ = 10, 30, 50) with $r$ = 1.4, 1.8, and 3.6 $a_0$ using density matrix renormalization group (DMRG) method  with the STO-6G basis set. Their results demonstrated that the H$_n$ band gaps decrease with increasing chain length $n$ for all three geometries, but due to the high computational cost of DMRG they were not able to determine the band gap at the thermodynamic limit. Here, we test performance of EA- and IP-ADC for simulating band gaps of H$_n$ ($n$ = 10 - 150) at $r$ = 1.4 $a_0$ (``compressed'') and 1.8 $a_0$ (``equilibrium'') with the STO-6G basis set against results from DMRG and EOM-CCSD.

\cref{fig:band_gap} shows band gaps computed using EA-/IP-ADC and EOM-CCSD for H$_n$ with $n$ = 10 - 150, as well as the reference values from DMRG for $n$ = 10 - 50. For short chain lengths ($n \le 50$), the best agreement with DMRG is shown by EOM-CCSD, which yields errors of less than 0.05 eV for $r$ = 1.4 $a_0$ and $\sim$ 0.25 eV for $r$ = 1.8 $a_0$. ADC(2) and ADC(3) show very similar results, significantly underestimating the band gap for both geometries. For H$_{50}$, the errors of both approximations are $\sim$ 0.6 and 1.4 eV for $r$ = 1.4 and 1.8 $a_0$, respectively. As the chain length increases, the computed band gaps gradually approach the thermodynamic limit. For the near-equilibrium $r$ = 1.8 $a_0$ geometry, the H$_{130}$ and H$_{150}$ band gaps only differ by $\sim$ 0.06 eV, indicating that they are nearly converged to the thermodynamic limit, which we estimate to be $\sim$ 4.3 and 3.1 eV for EOM-CCSD and ADC(2), respectively. At the compressed $r$ = 1.4 $a_0$ geometry, convergence of the computed band gaps with the chain length is slower. Assuming this convergence is monotonic, we estimate the $r$ = 1.4 $a_0$ thermodynamic limits for the EOM-CCSD and ADC(2) band gaps to be $\sim$ 3.35 and 2.8 eV. Increasing the perturbation order of the ADC approximations from ADC(2) to ADC(3) does not reduce the errors in the computed band gaps, indicating the slow convergence of the finite-order ADC perturbation expansion. This, coupled with the large difference between the EOM-CCSD and ADC results, suggests that the infinite-order electron correlation effects are very important for predicting accurate band gaps of the equally-spaced hydrogen chains. 

\section{Conclusions}
\label{sec:conclusions}
In this work, we presented an implementation of algebraic diagrammatic construction (ADC) theory for simulating electron attachment (EA) and ionization (IP) energies of molecules (EA-/IP-ADC). In our EA-/IP-ADC implementation, energies and intensities of the EA and IP transitions are computed from poles and residues of the one-particle Green's function that is approximated up to the third order in single-reference perturbation theory (EA- and IP-ADC(n), n = 2, 3). We presented two algorithms for solving the EA-/IP-ADC equations: (i) conventional algorithm that uses iterative diagonalization techniques to compute low-energy EA, IP, and density of states, and (ii) Green's function algorithm (GF-ADC) that solves a system of linear equations to compute density of states directly for a specified spectral region, providing access to EA and IP at high energies.

To assess accuracy of the EA-ADC(n) (n = 2, 3) methods, we benchmarked their performance for a set of atoms (He -- Kr), small molecules, and five DNA/RNA nucleobases. For all systems, our results demonstrate that the accuracy of EA-ADC(3) is comparable to that of equation-of-motion coupled cluster theory with single and double excitations (EA-EOM-CCSD), with mean absolute errors of $\sim$ 0.05 eV for closed-shell atoms and molecules and $\sim$ 0.15 eV for open-shell systems. Although both methods have the same $\mathcal{O}(N^6)$ computational scaling with the size of the basis set $N$, EA-ADC(3) is more efficient due to the non-iterative and Hermitian nature of its equations. EA-ADC(2) shows somewhat larger mean absolute errors of $\sim$ 0.1 eV and $\sim$ 0.2 eV for closed- and open-shell systems, respectively, but has an advantage of the lower $\mathcal{O}(N^5)$ computational scaling.

To demonstrate capabilities of the Green's function ADC algorithm, we employed the GF-IP-ADC implementation to compute core-level $K$-, $L$-, and $M$-shell ionization energies of a zinc atom without introducing core-valence separation (CVS) approximation, which is a widely used technique for calculating energies of core electrons. By comparing the ionization energies computed using GF-IP-ADC and CVS-approximated IP-ADC (CVS-IP-ADC), our results establish that the CVS approximation is very accurate for ionization of the $K$-shell ($1s$) electrons, but introduces significant errors (up to $\sim$ 11 eV) for the $L$ and $M$ electronic shells ($2s$, $2p$, $3s$, etc). Importantly, the CVS errors do not exhibit clear correlation with ionization energy or the order of the IP-ADC(n) method n, indicating an uncontrolled nature of the CVS approximation.

As our final test, we combined our implementations of EA- and IP-ADC(n) (n = 2, 3) to compute band gaps of the equally-spaced linear H$_n$ chains ($n$ = 10 - 150) at two internuclear distances ($r$ = 1.4 and 1.8 $a_0$) and compared them with reference results from EA-/IP-EOM-CCSD and density matrix renormalization group (DMRG). The computed band gaps decrease with increasing chain length $n$. At the near-equilibrium bond length $r$ = 1.8 $a_0$, the band gaps are nearly converged to the thermodynamic limit for $n$ = 150, while for the shorter $r$ = 1.4 the convergence with chain length is slower. For short chains, the EOM-CCSD band gaps are in a good agreement with DMRG, while both ADC(2) and ADC(3) methods significantly underestimate band gaps for all $n$. The errors in the ADC band gaps increase with $r$ and do not become smaller with increasing order of the ADC approximation, suggesting that infinite-order correlation effects are important for predicting accurate band gaps of hydrogen chains. 

Overall, our results demonstrate that EA- and IP-ADC are useful methods for accurate and efficient simulations of electron affinities, ionization potentials, and densities of states of weakly correlated chemical systems. To apply EA- and IP-ADC to larger and more complex systems, an efficient implementation of these methods is desirable. Another important direction is to improve accuracy of EA-/IP-ADC for systems with unpaired electrons and moderate or strong electron correlation, for example, by developing multi-reference formulations of these methods.\cite{Sokolov:2018p204113,Chatterjee:2019p5908} Work along these directions is ongoing in our group.

\section{Acknowledgements}
This work was supported by start-up funds provided by the Ohio State University. Computations were performed at the Ohio Supercomputer Center under projects PAS1317 and PAS1583.\cite{OhioSupercomputerCenter1987} 


\end{document}